\documentclass[3p,times,twocolumn]{elsarticle}

\usepackage{ecrc}

\volume{00}

\firstpage{1}

\journalname{Nuclear Physics B Proceedings Supplement}

\runauth{Thierry Lasserre}

\jid{nuphbp}

\jnltitlelogo{Nuclear Physics B Proceedings Supplement}

\usepackage{amssymb}

\usepackage[figuresright]{rotating}

\def\nuebar{\bar{\nu}_{\lowercase{e}}}

\begin{document}

\begin{frontmatter}



\dochead{}

\title{Testing the Reactor and Gallium Anomalies with Intense (Anti)Neutrino Emitters}

\author[label1,label2]{Thierry Lasserre}
\address[label1]{CEA, Irfu, SPP, Centre de Saclay, F-91191 Gif-sur-Yvette, France}
\address[label2]{Astroparticule et Cosmologie APC, 10 rue Alice Domon et L\'eonie Duquet, 75205 Paris cedex 13, France}

\begin{abstract}
Several observed anomalies in neutrino oscillation data could be explained 
by a hypothetical fourth neutrino separated 
from the three standard neutrinos by a squared mass difference 
of  a few 0.1 eV$^2$ or more. This hypothesis can be tested with MCi neutrino 
electron capture sources ($^{51}$Cr) or kCi antineutrino $\beta$-source ($^{144}$Ce) 
deployed inside or next to a large low background neutrino detector.  
In particular, the compact size of this source coupled with the localization of 
the interaction vertex lead to an oscillating pattern in event spatial 
(and possibly energy) distributions that would unambiguously determine neutrino 
mass differences and mixing angles.

\end{abstract}

\end{frontmatter}


\section{Introduction}
\label{sec:introduction}

Over the last 20 years neutrino oscillations associated with small splittings between 
the $\nu$ mass states have become well established. 
Three $\nu$ flavors ($\nu_e$, $\nu_{\mu}$, $\nu_{\tau}$) are mixtures of three massive 
neutrinos ($\nu_1$, $\nu_2$, $\nu_3$) separated by squared mass
differences of $\Delta m_{21}^2 = 8 \cdot 10^{-5}~{\rm eV}^2$ and 
$\Delta m_{31}^2 = 2.4  \cdot 10^{-3}~{\rm eV}^2$~\cite{pdg}.  
This is a minimal extension of the Standard Model that requires a lepton mixing matrix, 
similarly to the quark sector, and $\nu$ mass.
Beyond this model indications of oscillations between active and sterile $\nu$s have been observed 
in the LSND~\cite{LSND}, MiniBooNE~\cite{MiniBoone}, gallium~\cite{GA} and reactor~\cite{RAA} experiments.
This suggests the existence of a fourth massive $\nu$ with a mass of $\gtrsim$0.1~${\rm eV}^2$~\cite{SnuWP}. Testing $\nu$ anomalies now requires energy and baseline-dependent signatures for an unambiguous resolution. With MeV $\nu$'s very short baselines and compact sources become mandatory.

\section{Gallium and Reactor Anomalies}
\label{sec:raaga}

Man-made $\nu$ sources were originally proposed and built to measure the efficiency of solar-$\nu$ detection. 
First Alvarez proposed a $^{65}$Zn  source~ \cite{Alv73} producing 1.35 MeV neutrinos. 
Sources of lower energy neutrinos, more appropriate for the efficiency calibration of gallium based experiments 
were later proposed. Performing systematic searches for candidate nuclei using as criteria the activation cross section,
 isotopic abundance, absence of higher energy gamma-rays, and similarity of the $\nu$ spectrum to the solar 
 spectrum, two candidates were selected, $^{51}$Cr ($<$750~keV) proposed by Raghavan~\cite{Rag78} 
 and $^{37}$Ar (814~keV) proposed later by Haxton~\cite{Hax88}. 
In the nineties two $^{51}$Cr $\nu$-sources ($\mathcal{A}_0\sim$MCi) ) were made in the solar 
$\nu$ Gallex collaboration~\cite{Ans95} one in the Sage collaboration~\cite{Abd99} 
complemented with an $^{37}$Ar source ($\mathcal{A}_0$=0.4~MCi)~\cite{Abd06}. 
Both experiments observed an average deficit of \mbox{$R_G=0.86\pm 0.06\,(1\sigma)$}.
Fitting the data with the hypothesis of $\nu_e$ disappearance caused by
short baseline oscillations leads to \mbox{$|\Delta m_{\rm new}^2| > 0.3$~eV$^2$} (95\%) and  
\mbox{$\sin^2(2\theta_{\rm new}) \sim  0.2$}~\cite{GA,RAA}, 
assuming only one additional sterile $\nu$ (the so-called 3+1 model).

Recently the hypothetical existence of a fourth $\nu$ has been revived
by a new calculation~\cite{Mueller2011} of the rate of $\nuebar$ production
by nuclear reactors that yields a $\nu$ flux about 3.5\% higher than
previously predicted. Coupled with cross section reevaluations this result 
 implies that the measured event rates for all reactor $\nuebar$ experiments
within 100~meters of the reactor are 7\% too low, 
with an average deficit of \mbox{$R_R=0.927\pm 0.023\,(1\sigma)$}~\cite{RAA} .
The deficit could also be explained by a hypothetical fourth massive $\nu$
separated from the three others by $|\Delta m_{\rm new}^2|>0.1$~eV$^2$ 
and \mbox{$\sin^2(2\theta_{\rm new}) \sim  0.2$}~\cite{RAA}. 
  
The combination of the reactor $\nuebar$ anomaly with the gallium $\nu_e $ anomaly
disfavors the no-oscillation hypothesis at 99.9\%~C.L.~\cite{RAA,SnuWP}. 

\section{Searching for a $\Delta$ m$^2$ $>0.1$ eV$^2$ new $\nu$ state}
\label{sec:search4thnu}

Both reactor and gallium anomalies rely on the observation of the disappearance MeV-scale 
$\nuebar$'s and $\nu_e$'s by counting experiments. Therefore the definitive test of the anomalies 
is not only to test the $\nu$ disappearance at short baselines, but also to search for an oscillation
pattern as a function of L/E. 

Probing a $\Delta m^2$ of $\sim$1~eV$^2$ implies that an oscillation search using neutrinos with energies 
of typical of radioactive decays, in the few MeV range, requires a baseline of several meters only.
Therefore, assuming CP invariance, both anomalies could be unambiguously tested using $\nuebar$/$\nu_e$ 
emitters deployed at the center or next to a large detector, $\sim$10~m-scale, such as Sage-2Z~\cite{SnuWP}, 
Borexino~\cite{BorexDet}, KamLAND~\cite{KamLANDSol}, SNO+~\cite{SNO+}, or Daya Bay~\cite{DB+}.

\section{A new (anti)neutrino source experiment}
\label{sec:nusourceexp}

There are two suitable $\nu$-sources options for searching for light sterile neutrinos:
monochromatic $\nu_e$ emitters, like $^{51}$Cr or $^{37}$Ar, or
$\nuebar$ emitters with a continuous $\beta$-spectrum, like $^{144}$Ce or $^{106}$Ru.
In both cases the source must be as compact as possible to allow the observation of the characteristic 
$\nu$-oscillation pattern of event positions, even for $|\Delta m_{\rm new}^2|>$few~eV$^2$. 

\subsection{Neutrino Emitters}
\label{sec:nusource}

Radioactive neutrino sources involve either $\beta^+$-decay or electron capture.  
Electron capture decays produce mono-energetic  $\nu_e$'s  allowing for a determination 
of $L/E$ by measuring only the interaction vertex.
Intense man-made $\nu_e$ source were used for the calibration of
solar-$\nu$ experiments. In the nineties, $^{51}$Cr and $^{37}$Ar 
 were used as a check of the radiochemical
experiments Gallex and Sage~\cite{SolCalib90s}.
Production of an $^{37}$Ar source requires a large fast breeder reactor 
that leaves $^{51}$Cr as the best current  $\nu_e$  source candidate 
for sterile $\nu$ search.  

$^{51}$Cr has a half-life of 27.7~days.  90.1\% of the time it decays to the ground state of $^{51}$V 
and emits a 751~keV $\nu_e$ while 9.9\% of the time it decays to the first excited state of $^{51}$V 
and emits a 413~keV $\nu_e$ followed by 320~keV $\gamma$.  $^{50}$Cr has a relatively high average 
thermal neutron capture cross section of 17.9~barn that makes the large scale production of $^{51}$Cr possible.  
Natural Cr is primarily $^{52}$Cr (83.8\%) and  contains 4.35\% $^{50}$Cr.  The isotope $^{53}$Cr 
(9.5\% of natural chromium) has an average thermal neutron cross section of 18.7~barn, 
so when natural chromium is irradiated, $^{53}$Cr absorbs 2.5 neutrons to every one captured on $^{50}$Cr, 
reducing the $^{51}$Cr yield. Therefore enriched $^{50}$Cr is needed for reaching several MCi of activity.
Enrichment would also play in favor of manufacturing a compact target necessary for the 
sterile neutrino search. The material used by Gallex was enriched to 38.6\% in $^{50}$Cr while the Sage 
target was enriched to 92\%.  
Because many isotopes have high neutron capture cross sections great care must 
be taken during the production and handling of the Chromium rods to minimize the 
introduction of chemical impurities leading to high-energy gamma rays.

In radiochemical experiments the interaction of  $^{51}$Cr and $^{37}$Ar
neutrinos induce the reaction $^{71}$Ga($\nu_e$,e$^-$)$^{71}$Ge. 
The $^{71}$Ge produced is chemically extracted from the detector and 
converted to GeH$_4$. Ge atoms are then placed in proportional counters 
and their number is determined by counting the Auger electrons released 
in the transition back to $^{71}$Ga, which occurs with a half life of 11.4~days. 

In LS experiment the signature is provided by $\nu_e$ elastic
scattering off electrons (ES) in the LS molecules. The cross section is 
$\sigma(E_\nu) \sim 0.95 \,  10^{-44}\times E_\nu \,\, \rm{cm}^2$, 
where  $E_\nu$ is the neutrino energy in MeV. This signature
can be mimicked by Compton scattering induced by radioactive and
cosmogenic background, or by Solar-$\nu$  interactions. The
constraints of an experiment with $\nu_e$ impose the use of
a very high activity source, 5-10 MCi, to provide a production
rate in the detector that will exceed the rate from the Sun, and to
compensate the loss of solid angle due to the location of the source 
outside of the detector (since the ES does not 
provide any specific signature of  $\nu_e$ interaction).

\subsection{Antineutrino Emitters}
\label{sec:antinusource}

Antineutrino sources are non-monochromatic $\nuebar$ emitters decaying through $\beta$-decay. 
$\beta^-$-decay induced $\nuebar$ are detected through the inverse beta-decay (IBD) reaction
$\bar{\nu}_e$ + p $\rightarrow$ $e^+$+n. 
The IDB cross section is $\sigma(E_e) \sim 0.96 \,  10^{-43}\times p_e E_e \,\,
\rm{cm}^2$, where $p_e$ and $E_e$ are the momentum
and energy (MeV) of the detected $e^+$, neglecting
recoil, weak magnetism, and radiative second order corrections. This lead to an interaction rate of 
an order of magnitude higher than the EC process at 1 MeV, allowing to reduce the activity 
to the kCi scale for the sterile neutrino search~\cite{CeLAND}

The $e^+$ promptly deposits its kinetic energy in the LS and annihilates emitting 
two 511~keV $\gamma$-rays, yielding a prompt event, with a visible energy of E$_e$=
E$_\nu$-($m_n$-$m_p$)~MeV; the emitted keV neutron is captured on a free proton
with a mean time of a few hundred microseconds, followed by the emission of a
2.2~MeV deexcitation $\gamma$-ray providing a delayed coincidence event. 
The delayed coincidence between detection of the positron and the neutron capture 
gamma rays suppressed any non-source background to a negligible level.

A suitable $\nuebar$ source must have $Q_\beta>$1.8~MeV (the reaction
threshold) and a lifetime that is long enough ($\gtrsim$~6~months) to allow for
production, transportation, and deployment in the detector. For individual nuclei,
these two requirements are contradictory so one expect candidate
sources to involve a long-lived low-$Q$ nucleus that decays to a short-lived
high-$Q$ nucleus. Four such pairs have been identified~\cite{Kor94,CeLAND}: 
$^{144}$Ce-$^{144}$Pr  ($Q_\beta$(Pr)=2.99~MeV), 
$^{106}$Ru-$^{106}$Rh ($Q_\beta$(Rh)=3.54~MeV), 
$^{90}$Sr-$^{90}$Y ($Q_\beta$(Y)=2.28~MeV), 
and $^{42}$Ar-$^{42}$K  ($Q_\beta$ (K)=3.52~MeV).
The first three are common fission products from nuclear reactors that
can be extracted from spent fuel rods.

\section{Current proposals}
\label{sec:proposals}

We present here a non-comprehensive description of possible future neutrino source 
experiments dedicated to sterile neutrino oscillation search. 
The projects described below are in various stages of development from early stage to conceptual.  

\subsection{$^{51}$Cr neutrino projects}
\label{sec:51Cr}

\begin{figure}
\begin{center}
	\resizebox{\linewidth}{!}{\includegraphics{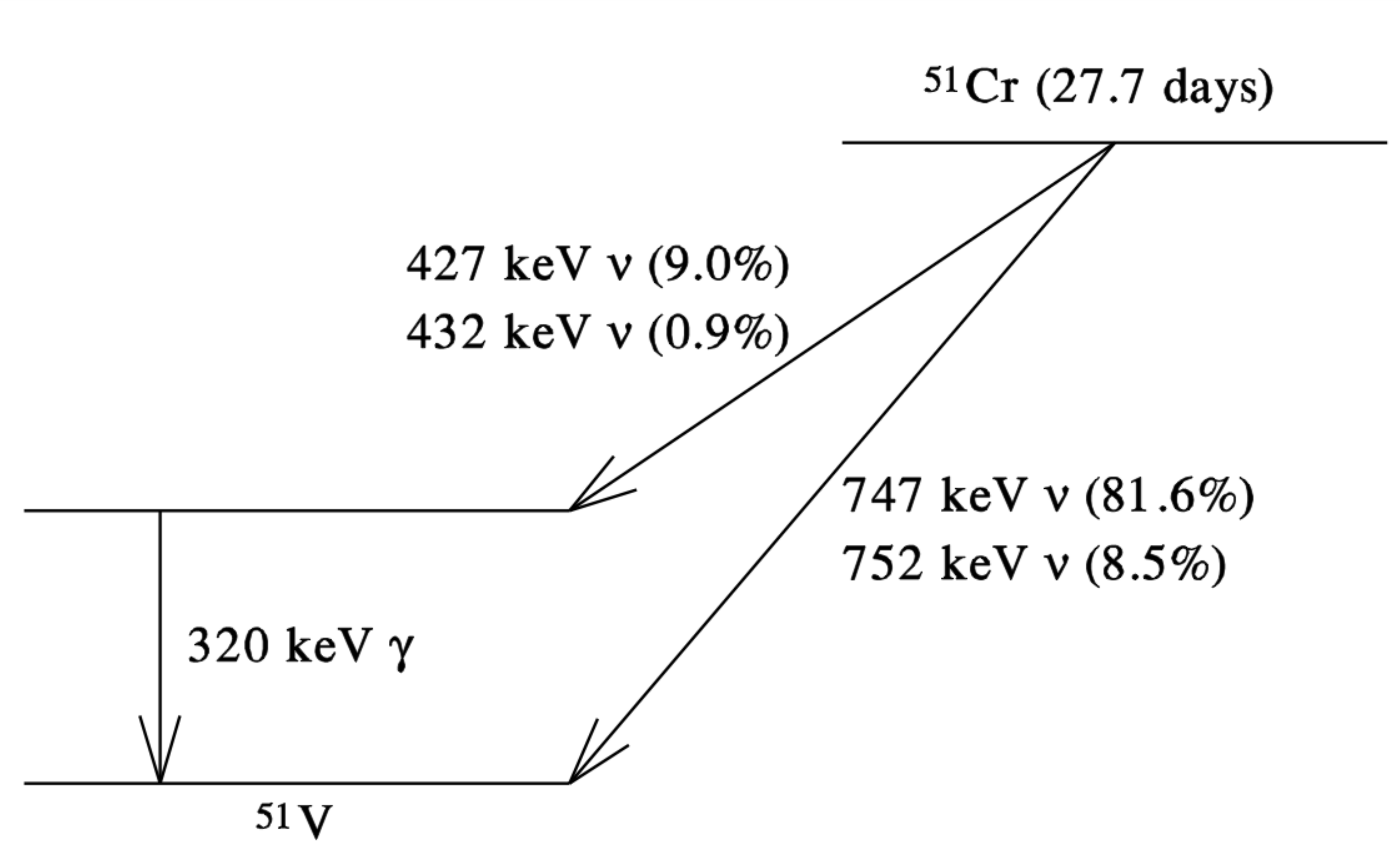}}
	\caption{Decay scheme of $^{51}$Cr to$^{51}$V through electron capture.}
	\label{fig:cr51}
\end{center}
\end{figure}

We first review projects based on $^{51}$Cr $\nu_e$ emitters whose 
decay scheme is shown on Fig~\ref{fig:cr51}.

\subsubsection{Dual Metallic Gallium Target at Baksan}
\label{sec:sage}

The goal is to place a $^{51}$Cr source with initial activity of 3~MCi at the center
 of a 50-ton target of liquid Ga metal. The target will be divided into two
concentric spherical zones, an inner 8-ton zone and an outer 42-ton zone. 
In this two-zone experiment the source dimension will be on the scale of 10~cm 
and the baseline will be one the scale of a few~meters, therefore the oscillation  
ripples won't be averaged out for $\Delta m^2 \approx1$~eV$^2$.
65~atoms of $^{71}$Ge are expected each day in each zone at the beginning 
of each run. A sequence of exposure period of a few days will be done. 
The extraction and the counting will be the same as for the well
understood SAGE experiment. Assuming 10 extractions, each with a 9-day 
exposure, one expects a total uncertainty of $\pm4.5$\%. 
The results will partially rely on the knowledge of the activity of the source
that will be measured by calorimetry (a 3~MCi source releases initially 650~W of heat). 
Either a significant difference between the capture rates in the two zones, or an 
average rate in both zones significantly below the expected rate would be an
evidence of nonstandard neutrino properties.
The proposed experiment has the potential to test neutrino
oscillation transitions with mass-squared difference $\Delta
m^2>0.5$~eV$^2$ and mixing angle $\theta$ such that $\sin^2 2\theta > 0.1$.  
To conclude this project relies on a proofed concept, e.g., the measurement of the
solar neutrino flux by SAGE for many years, with well understood backgrounds and 
systematics. A relevant aspect concerns the accompanying gamma radiation
related to impurities which are not able to produce the reaction 
$^{71}$Ga($\nu_e$,e$^-$)$^{71}$Ge. A high activity a $^{51}$Cr source can 
thus be deployed inside such a neutrino detector and the requested shield 
surrounding the source is mainly used to fulfill safety requirements. 
 
\subsubsection{SOX and SNO+Cr}
\label{sec:sox}

After more than five years of data taking the solar-$\nu$ detector Borexino is 
well suited to host an external neutrino source experiment~\cite{Ian99}. 
A tunnel exists right below the water tank, providing a location at a distance 
of 8.25~m to the scintillator inner vessel center. 
The unique extreme radiopurity achieved in the LS medium will allows to control
 the irreducible contribution of $^7$Be solar neutrinos. The experiment will 
 consist in counting the number of observed events at each detector 
location and to compare it to the expectation without oscillations. The position of 
each event can be reconstructed with a precision of $\sim$12~cm at 1~MeV.
In order to highlight the physics reach of a test accomplished with an external 
source, Fig.~\ref{fig:sensitivities} displays the exclusion plot that might be 
obtained at 95\%~C.L., with a 10~MCi $^{51}$Cr source located externally.
Though not conclusive, an external test performed with a sufficiently strong neutrino
source will start to address a sizable portion of the oscillation parameter region 
of the Gallium and reactor anomalies. 

A similar program as been proposed in the SNO+ detector but deploying the source 
at the center of the detector~\cite{SnuWP}.
Since the detection will be performed via elastic scattering on electrons, 
the usual Compton effect  induced by gammas is a dangerous background. 
Feedback from Gallex experiment activity measurement performed after the 
preparation $^{51}$Cr source indicates a rate of 10~GBq of long lived 
1.5 MeV $\gamma$'s~\cite{SolCalib90s}. Those $\gamma$'s must thus be shielded
via a thick dense alloy like tungsten, that also needs to be ultrapure
at the mBq/kg level.

We mention here that the Borexino contemplates first the deployment a $^{51}$Cr 
$\nu_e$ external source and second the deployment of an internal $\nuebar$ 
source,  like the $^{144}$Ce-$^{144}$Pr material (see Section \ref{sec:celand}).

\subsubsection{LENS-Sterile}
\label{sec:sno}

The LENS (for Low Energy Neutrino Spectroscopy) project is first intended to 
measure solar neutrinos energy in real time~\cite{Lens}. The LENS-Sterile concept 
consist in placing a 10 MCi neutrino source at the center of detector, and counting 
the $\nu_e$ interactions as a function of distance from the source~\cite{Lens+}. 
The detection will be done through a low threshold charged current process,  
$^{115}{\rm{In}}$ + $\nu_e$  $\rightarrow$  $^{115}{\rm{Sn}}^*$ + $e^-$, followed
 by gamma ray's de-exitation allowing to reject backgrounds. 
Beyond the very high activity source production the first challenge consist in the 
realization of a low background segmented neutrino detector doped with natural indium.

\subsection{$^{144}$Ce-$^{144}$Pr projects}
\label{sec:144Ce}

\begin{figure}
\begin{center}
	\resizebox{\linewidth}{!}{\includegraphics{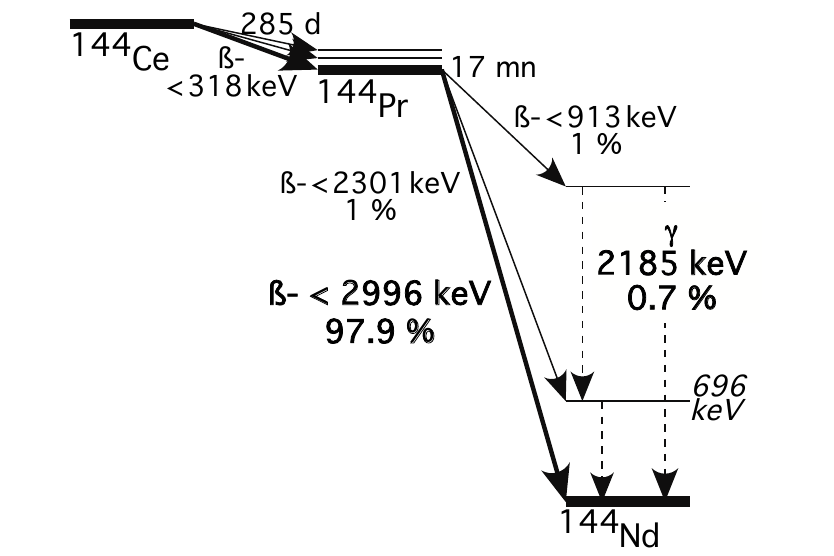}}
	\caption{In this simplify decay scheme the relatively long lifetime $^{144}$Ce 
	decay into the short life $^{144}$Pr decaying itself mainly through an energetic 
	$\beta$ extending up to 2996 keV.}
	\label{fig:ce144}
\end{center}
\end{figure}

We now review projects based on $^{144}$Ce-$^{144}$Pr $\nuebar$ emitters whose 
decay scheme is shown on Fig~\ref{fig:ce144}.

\subsubsection{CeLAND}
\label{sec:celand}

\begin{figure}
\begin{center}
	\resizebox{\linewidth}{!}{\includegraphics{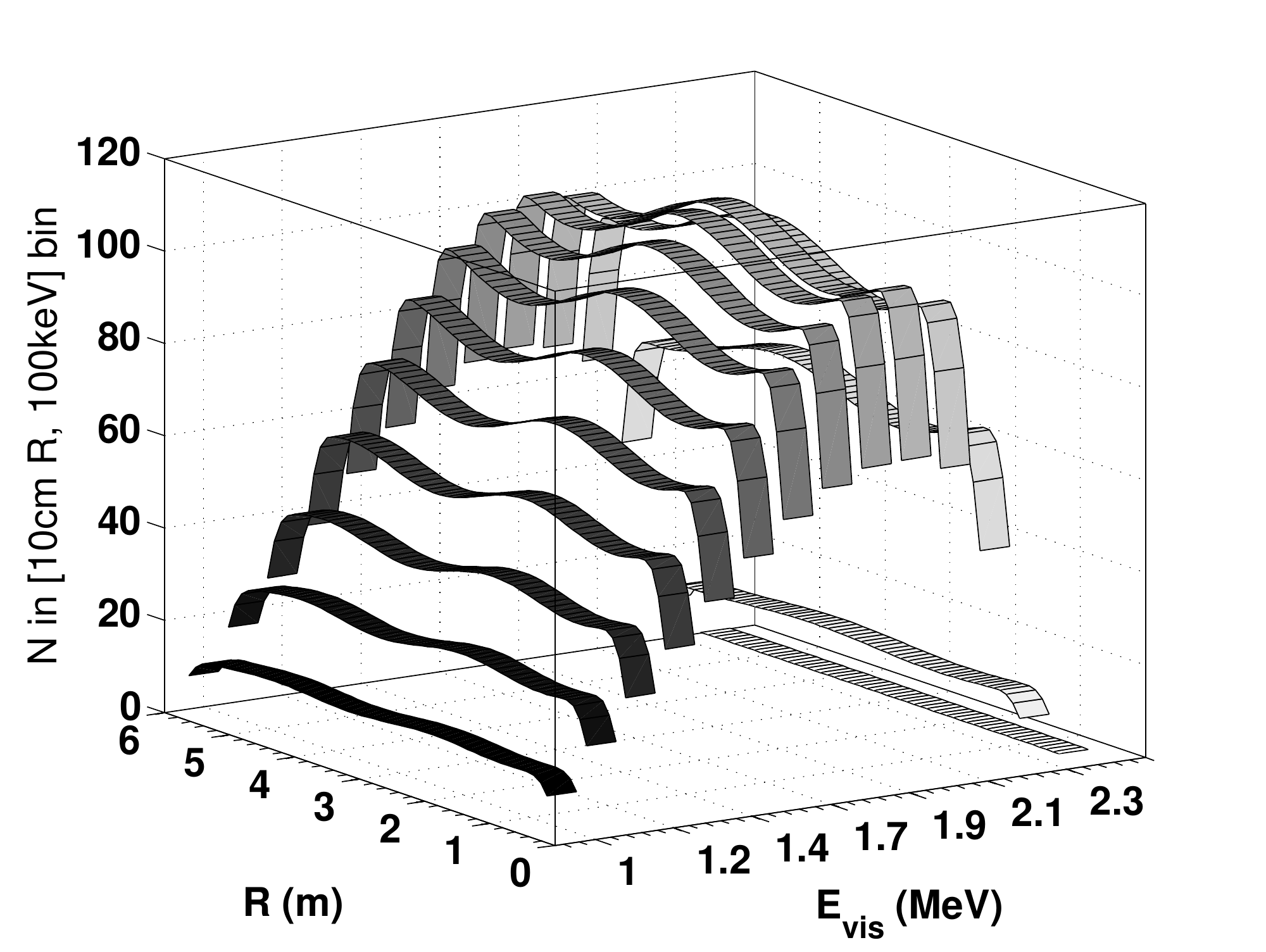}}
	\caption{Signal of a $^{144}$Ce source deployed at a center of a
	 LS detector, providing both R and E$_{vis}$ oscillation patterns
	 (E$_{vis}$=E$_e$+2m$_e$).}
	\label{fig:ce144_signal}
\end{center}
\end{figure}

\begin{figure}
\begin{center}
	\resizebox{\linewidth}{!}{\includegraphics{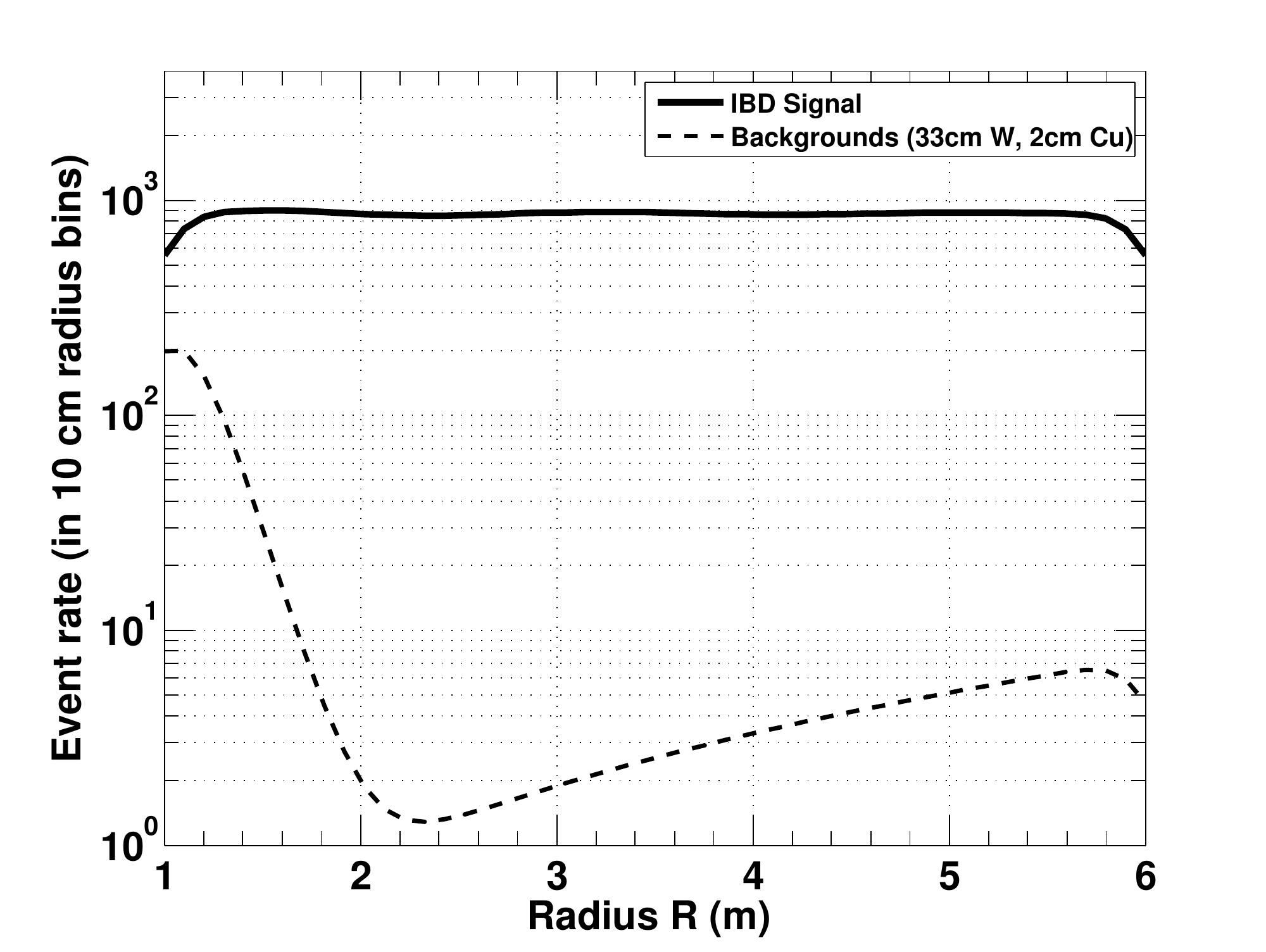}}
	\caption{Signal rate of a 50~kCi $^{144}$Ce deployed for 1 y 
	at a center of a LS detector (plain line), compared to the sum of 
	 backgrounds (dashed line), as a function of the detector radius 
	 in 10 cm concentric bins.}
	\label{fig:celand_sb}
\end{center}
\end{figure}

CeLAND is a project based on 50 kCi of $^{144}$Ce~\cite{CeLAND}. Cerium was 
chosen because of its high $Q_\beta$, its $\sim$4\% abundance in fission products 
of  uranium and plutonium, and finally for engineering considerations related 
to its possible extraction of rare earth from regular spent nuclear fuel reprocessing 
followed by a customized column chromatography. 
While not minimizing the difficulty of doing this, the nuclear industry does have the
technology to produce sources of the appropriate intensity, at the ppm purity level 
and first samples are in the processed of being delivered in 2013.
The goal of CeLAND is to deploy the $^{144}$Ce radioisotope at the center or next to 
 a large LS detector, like KamLAND, Borexino, or SNO+,  and to search for an oscillating 
 pattern in both event spatial and energy distributions that would determine 
neutrino mass differences and mixing angles through an unambiguously L/E signature.
We now focus on the unique oscillation signature induced by an eV-scale sterile $\nu$ 
at the center of a large neutrino detector. 
For $^{144}$Ce-$^{144}$Pr, 1.85 PBq (50 kCi) source lead to 40,000 interactions in one 
year in a KamLAND-like detector, between 1.5 and 6~m away from the source.
This is realized with $\sim$15~g of $^{144}$Ce, whereas the total mass of all
cerium isotopes is a few kg, for an extraction from selected fission products, 
as fresh as possible (e.g. 2 to 3 years after the last irradiation).
Thanks to cold-pressing technics the source fits inside a $<$5~cm-scale capsule,  
small enough to consider the Cerium ball as a point-like source. 
For comparison the vertex reconstruction is $<$15 cm.  
$^{144}$Ce has a low production rate of high-energy $\gamma$ rays ($>1MeV$) 
from which the $\nuebar$ detector must be shielded to limit background events.  
This source initially releases $\sim$300 W, but it will be self-cooled through 
convective exchanges with the LS without increasing the detector temperature
 by more than a few degrees. 
 The expected oscillation signal for $\Delta m_{\rm new}^2=2$~eV$^2$
and $\sin^2(2\theta_{\rm new})=0.1$, is shown on Fig.~\ref{fig:ce144_signal}. 
The space-time coincidence signature of IBD events ensure an almost background-free 
detection. Backgrounds are of two types, those induced by the environment or detector, 
and those due to the source and its shielding. 
 The main concern is accidental coincidences between a prompt
(E$>$0.9 MeV) and a delayed energy depositions (E$=$2.2 MeV) occurring within a
time window taken as three neutron capture lifetimes on hydrogen 
($\sim$600$\mu$sec), and within a volume of 20 m$^3$
The main source of detector backgrounds originates from
accidental coincidences, fast neutrons, and the long-lived muon induced
isotopes $^9$Li/$^8$He and scales with $R^2$ when using concentric $R$-bins. 
These components are routinely being measured in-situ in Borexino and KamLAND.
Geologic $\nuebar$ arising from the decay of radioactive isotopes of
Uranium/Thorium in the Earth have been measured 
in KamLAND~\cite{KamLANDGeo} and Borexino~\cite{BorexGeo}. 
Reactor $\nuebar$ emitted by the $\beta$-decays of the fission
products in the nuclear cores have been measured in KamLAND~\cite{KamLANDGeo}. 
The sum of all these backgrounds is quite small with
respect to the $\nuebar$ rate from a kCi source. Note that non-source backgrounds
 can be measured in-situ during a blank run with an empty shielding.
The most dangerous source background originates from the energetic
2.185~MeV $\gamma$ produced by the decay through excited states of $^{144}$Pr.  
We approximate $\gamma$ ray attenuation in a shield of~$\sim$35~cm of tungsten 
alloy with an exponential attenuation law accounting for Compton scattering 
and photoelectric effect. The intensity these $\gamma$ rays is then decreased by 
a factor $<10^{-12}$~\cite{Nist}, to reach a tolerable rate. 
An important remaining background source could be the tungsten alloy shield
itself. Activities at the level of ten to hundreds mBq/kg have been
reported, suitable for the experiment.  
Assuming a $\sim$5 tons shield a prompt and delayed event
 rates of 50~Hz and 25~Hz, respectively source induced background  become 
negligible beyond a distance of 1.5~m from the source. An oil buffer surrounding the 
shielding or a composite shielding could further suppress both source and shield 
backgrounds if necessary; Any of the photons or shielding backgrounds can
account for either the prompt or delayed event, depending on their energy. 
The sum of the backgrounds integrated over their energy spectrum is
shown on Fig.~\ref{fig:celand_sb}, supporting the case of kCi $\nuebar$
source versus MCi $\nu_e$ source for which solar-$\nu$'s become an
irreducible background.
Assuming a new neutrino oscillation with $\Delta m_{\rm new}^2=2$~eV$^2$ and 
$\sin^2(2\theta _{\rm new})=0.1$, the interaction rate decreases from 40,000 to 
38,000 per year. The sensitivity of an experiment with a 50~kCi $^{144}$Ce source 
running for 1~year, using only events between 1.5~m and 6~m is displayed on 
Fig~\ref{fig:sensitivities}. The 95\% C.L. sensitivity is extracted through a 
Pearson $\chi^2$ test assuming a 2\% fully uncorrelated systematic error, and
 accounting for a fiducial volume uncertainty of 1\% in a calibrated
detector, as well as for ($e^+$, n)  space-time coincidence detection
efficiencies uncertainties at the sub-percent level. 
The source activity uncertainty is taken as a normalization error of 1\%
Results indicates that 50~kCi of $^{144}$Ce allows us to probe the whole combined 
reactor and Gallium anomaly parameter space at least at 95\% C.L. An analysis 
assuming no knowledge on the source activity shows that the oscillatory behavior 
can be established  for $\Delta m_{\rm new}^2 <10$~eV$^2$.

\subsubsection{$^{144}$Ce-$^{144}$Pr in Daya-Bay}
\label{sec:db}

Following the original CeLAND proposal~\cite{CeLAND} the far site detector complex of the 
Daya Bay reactor experiment is being considered for the deployment of a 500 kCi
 $^{144}$Ce-$^{144}$Pr towards the search of sterile neutrinos~\cite{DB+}. 
The far site detector complex of the Daya Bay reactor experiment houses four 20-ton 
$\nuebar$ detectors with a separation of 6~m. 
When combined with a compact radioactive $\nuebar$ source the detectors provide a 
well suited setup for the study of short baseline neutrino oscillation with multiple detectors 
over baselines ranging from 1.5-8~m. The source would be deployed in the water pool 
surrounding the four far-site detectors providing natural shielding minimizing technical 
complications. However a 18.5 PBq $^{144}$Ce (16~cm in diameter) is necessary
to reach enough statistics. Contrarily to other proposals the dominant background will come
 from reactor $\nuebar$'s.  However this can be measured to a few percent precision 
 through blank runs currently performed for the standard oscillation progam being carried out.
 As for CeLAND, the Daya Bay setup can probe sterile neutrino oscillations powerfully by measuring 
both spectral distortions of the energy and baseline spectrum.  Coupled to a well-measured
source rate normalization further information can be provided to test the anomalies.  

The proposed Daya Bay sterile neutrino experiment can probe the 0.3-10 eV$^2$ mass splitting 
range to a sensitivity of as low as sin$^2$2$\theta_{new}<$0.04 at 95\% CL.  The experiment will 
be sensitive to most of the 95\% CL allowed sterile neutrino parameter space suggested by 
the reactor and the Gallium anomalies (see Fig~\ref{fig:sensitivities}).  

\subsection{Conclusion}
\label{sec:conclusion}

\begin{figure}
\begin{center}
	\resizebox{\linewidth}{!}{\includegraphics{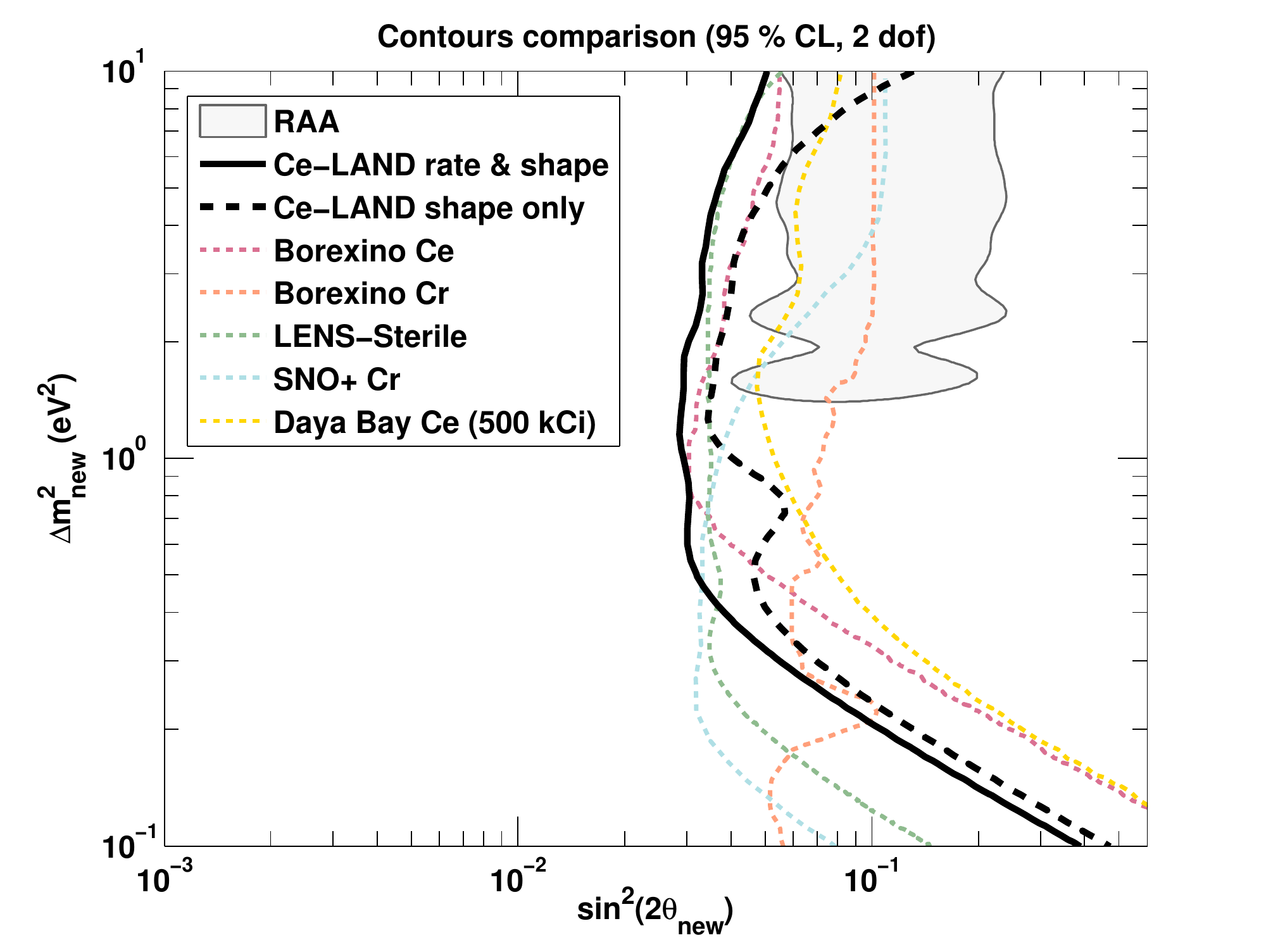}}
	\caption{95\% C.L. exclusion limit of the $\nu$ source projects planned 
	towards the search of a fourth neutrino state,
	in the $\Delta m_{\rm new}^2$ and $\sin^2(2\theta_{\rm new})$ plane (2~dof). 
	The best sensitivity is provided by the 50 kCi.y $^{144}$Ce project (CeLAND, thick black curves).
	Results are compared to the 95\% C.L. inclusion domains given by 
	the combination of reactor neutrino experiments, Gallex and Sage calibration 
	sources experiments, as described in Fig. 8 of~\cite{RAA}  (gray area).}
	\label{fig:sensitivities}
\end{center}
\end{figure}

The implications of the existence of an additional sterile neutrino states would be 
profound. This would require the evolution of the current adopted paradigm of the 
standard model of Particle Physics. As a result, great interest has developed in testing 
the hypothesis of sterile neutrinos. Novel $\nu$ source experiments could play a major 
role in order to provide a definitive resolution of the current neutrino anomalies. 
The global superior characteristic of the $\nuebar$ scenario is illustrated on sensitivity 
curves in fig. \ref{fig:sensitivities}, where the 95\% sensitivity contour covers the entire 95\% joint 
gallium-reactor anomaly region. At this early stage  both MCi $^{51}$Cr $\nu_e$ and $^{144}$Ce 
kCi $\nuebar$ sources projects are currently being developed, each facing the challenges 
of the source production, transportation, and deployment. In the forthcoming years these 
novel efforts could disprove or confirm convincingly the sterile neutrino hypothesis~\cite{SnuWP}.


\end{document}